# A meta-analysis of state-of-the-art electoral prediction from Twitter data

DANIEL GAYO-AVELLO, University of Oviedo (Spain)

Electoral prediction from Twitter data is an appealing research topic. It seems relatively straightforward and the prevailing view is overly optimistic. This is problematic because while simple approaches are assumed to be good enough, core problems are not addressed. Thus, this paper aims to (1) provide a balanced and critical review of the state of the art; (2) cast light on the presume predictive power of Twitter data; and (3) depict a roadmap to push forward the field. Hence, a scheme to characterize Twitter prediction methods is proposed. It covers every aspect from data collection to performance evaluation, through data processing and vote inference. Using that scheme, prior research is analyzed and organized to explain the main approaches taken up to date but also their weaknesses. This is the first meta-analysis of the whole body of research regarding electoral prediction from Twitter data. It reveals that its presumed predictive power regarding electoral prediction has been rather exaggerated: although social media may provide a glimpse on electoral outcomes current research does not provide strong evidence to support it can replace traditional polls. Finally, future lines of research along with a set of requirements they must fulfill are provided.

## 1 INTRODUCTION

Real world events have an impact in online systems and trails left by users on such systems have been used to perform early event detection in an automatic fashion. Indeed, a number of well-known studies have shown that flu [Ginsberg et al. 2009], unemployment rates [Choi and Varian 2009a], or car sales [Choi and Varian 2009b] can be forecasted on the basis of Web search queries.

Previously, but in a similar way, blog posts were correlated with book sales [Gruhl et al. 2005] or movie gross incomes [Mishne and Glance 2006]. It must be noted that while queries are not amenable to sentiment analysis, blog posts are and, hence, applying opinion mining to blogs stronger correlations can be found –e.g. [Mishne and de Rijke 2006; Mishne and Glance 2006].

When compared to blogs, microblogging is a recent phenomenon; nevertheless, it has attracted a good deal of attention on the basis that it can help to "take the pulse" of society. Indeed, the brevity of microposts coupled with the fact that a single service (i.e. Twitter) is the de facto standard for such kind of publishing, has turned this data into a new favorite for researchers.

Therefore, based on the data abundance offered by Twitter, and in light of the excellent results obtained in the past by mining query logs and blogs, it is unsurprising that a number of studies have been conducted on predicting both present and future events from tweets.

For instance, [Asur and Huberman 2010] exploited Twitter data to predict box-office revenues for movies; [O'Connor et al. 2010] rather successfully correlated tweets with several public opinion time series; [Tumasjan et al. 2010] claimed to have predicted the outcome of German elections, [Bollen et al. 2011] the stock market, and [Lampos and Cristianini 2010] the evolution of flu pandemics.

A superficial glimpse on that growing body of literature could suggest that Twitter data has an impressive –and versatile– predictive power. However, this has been questioned. For instance, [Wong et al. 2012] raised some doubts on the predictability of box-office performance from Twitter data; and, in a similar way, [Jungherr et al. 2011] rebutted the main thesis by Tumasjan et al., namely, that tweets and votes are strongly correlated.

Therefore, who is right? Those who claim Twitter data is a good predictor or their critics?

First of all, we must acknowledge a strong bias permeating research: The tendency of researchers to report positive results while not reporting negative ones. This bias, the so-called "file drawer" effect [Fanelli 2010], is extremely harmful.

Firstly, it makes us to assume that published (positive) results are the norm and, hence, that they are to be expected in the future.

Secondly, it makes difficult to publish negative results since the burden of proof lies on those criticizing the prevailing (positive) point of view.

That said it goes without saying that it is far from trivial proving or disproving most of the aforementioned methods of prediction. For one reason, most of them –namely, those involving box-office results, the stock market, epidemics, or opinion polls– are not predicting an event but showing correlation between two different time series: One produced from Twitter data and another obtained from the "off-line" world.

In that regard, such methods are models and, therefore, they can be more or less accurate and, in turn, more or less useful.

Nevertheless, there is an area where discrete results at specific moment –i.e. predictions– are expected: Politics.

Clearly, voters can be more or less close to a candidate but, eventually, each of them has to cast a vote. Similarly, polls change over time but once polling stations are closed, the number of ballots is fixed, and there is a winner.

As it has been implied above, there are opposing points of view with regards to the predictive power of Twitter data on elections. However, there is no controversy here: by analyzing prior reports we can reach sound conclusions on the presumed power of state-of-the-art methods and, much more important, we can:

1. Point out the main weakness in current approaches.
2. Suggest important challenges that are still open and core lines of future research; and
3. Provide recommendations and conclusions for those interested in pushing forward this area.

Such contributions require a thorough review of the body of literature. Nevertheless, more important than providing a bibliography it is to organize such different reports within a coherent conceptual scheme. That scheme is covered in the following section.

## 2 CHARACTERIZATION OF TWITTER-BASED ELECTORAL PREDICTION METHODS

Although unstated, it is assumed that any method to predict electoral results from Twitter data is an algorithm; otherwise, it would be impractical and pointless.

Therefore, such methods process some collection of tweets to make predictions; they are parameterized to adapt to different scenarios; and, finally, predictions can be more or less detailed (for instance, just providing the winner or vote rates for the different candidates) and they should be eventually evaluated against the actual results.

Thus, there are a number of characteristics and sub-characteristics defining any method to predict electoral results from Twitter; namely:

1. Period and method of collection: i.e., the dates when tweets were collected, and the parameterization used to collect them.
2. Data cleansing measures:
   a. Purity: i.e., to guarantee that only tweets from prospective voters are used to make the prediction.
   b. Debiasing: i.e., to guarantee that any demographic bias in the Twitter user base is removed.
   c. Denoising: i.e., to remove tweets not dealing with voter opinions (e.g. spam or disinformation) or even users not corresponding to actual prospective voters (e.g. spammers, robots, or propagandists).
3. Prediction method and its nature:
   a. The method to infer voting intentions from tweets.
   b. The nature of the inference: i.e., whether the method predicts individual votes or aggregated vote rates.
   c. The nature of the prediction: i.e., whether the method predicts just a winner or vote rates for each candidate.
   d. Granularity: i.e., the level at which the prediction is made (e.g. district, state, or national).
4. Performance evaluation: i.e., the way in which the prediction is compared with the actual outcome of the election.

The papers analyzed in this work were characterized according to that scheme. Table II shows that characterization and it also includes descriptive information for each of the elections.

## 2.1 General characteristics of research conducted up to date

The first thing that Table II reveals is that, surprisingly, literature regarding electoral prediction is not prediction at all. All of the reports were written post facto and, hence, those studies discussing promising or even positive results describe, in fact, how elections could have been predicted.

Moreover, although there is replication of results it is not usually conducted by the same authors proposing the original method. Fortunately, as it will be shown, only two "flavors" of voting inference in Twitter have been widely used –namely, tweet counts and lexicon-based sentiment analysis– and, thus, a number of papers have evaluated and compared both.

In addition to that, a number of papers deal with the same elections. [O'Connor et al. 2010] and [Gayo-Avello 2011] covered the US presidential election in 2008. And [Tumasjan et al. 2010] and [Jungherr et al. 2011] covered the German federal election in 2009. Those papers should clarify if consistent results can be obtained from Twitter data or if, by the contrary, results strongly depend on decisions made by the researchers.

The rest of papers correspond to single case scenarios but [Metaxas et al. 2011] deserves special attention since it covers six different races from the same elections in the United States. That paper casts light on whether positive results achieved by those methods can be due to pure chance.

With regards to the elections covered by the literature, the United States is the best studied scenario (4 papers), followed by Germany (2 papers); Ireland, Singapore and Netherlands are covered by one paper each.

## 2.2 Period of collection

Table II reveals substantial variation among studies with regards to some of the characteristics in the scheme.

For instance, the period of data collection varies widely. Table I shows that some studies collected data just one week before elections while others collected data for weeks, months or even years.

There is, however, consensus about the ending point for the period of collection: the day before elections. The only paper contravening this rule is [Tumasjan et al. 2010] and, as it will be shown, this was a matter of criticism [Jungherr et al. 2011].

It must be noted that it is unclear the impact that the period of collection has in the predictions. [Jungherr et al. 2011] showed that by using different time windows performance underwent substantial variations. [Metaxas et al. 2011], using just one week of data, were able to obtain both correct and incorrect predictions.

Hence, it is clear that further research is needed in this regard and compelling arguments are needed to choose a given period of collection.

Table I. Different periods of data collection used in the literature ordered by decreasing time length. All of the studies, except for the one by Tumasjan et al. finished data collection the day before elections.

| Authors | Start of collection | End of collection |
|---|---|---|
| Livne et al. 2011 | 3 years before election | election day |
| O'Connor et al. 2010 | 10 months before election | election day |
| Gayo-Avello 2011 | 5 months before election (candidate nomination) | election day |
| Tumasjan et al. 2010 | 7 weeks before election | one week before election day |
| Jungherr et al. 2011 | 7 weeks before election (to replicate findings by Tumasjan et al.) | election day |
| Skoric et al. 2012 | 1 month before election | election day |
| Bermingham & Smeaton 2011 | 3 weeks before election | election day |
| Metaxas et al. 2011 | 1 week before election | election day |
| Tjong Kim Sang & Bos 2012 | 1 week before election | election day |

| Authors | Election | Period & Method of collection | Data cleansing | Prediction method | Performance evaluation | Reported results |
|---|---|---|---|---|---|---|
| O'Connor et al. 2010 | US presidential election, 2008 (Nov. 4, 2008) | February to November 2008. Candidate names used as keywords. | No cleansing at all. | Lexicon-based sentiment analysis. Aggregated results at national level. No prediction attempted. | Correlation against pre-electoral polls. | No significant correlation found. |
| Gayo-Avello 2011 | | June 1 to November 3, 2008. Presidential and vice-presidential candidate names. | Geolocated tweets at county level. Attempt to debias data according to user age. | Lexicon-based sentiment analysis. Individual votes. Aggregated results at state level. Vote rates. | MAE against actual electoral results. | MAE 13.10% (uncompetitive with traditional polls). |
| Tumasjan et al. 2010 | German federal election, 2009 (Sept. 27, 2009) | August 13 to September 19, 2009. Parties present in the *Bundestag* and politicians from those parties. | No cleansing at all. | Number of tweets. Aggregated results at national level. Vote rates. | | MAE 1.65% (comparable with traditional polls, although larger). |
| Jungherr et al. 2011 | | Different time windows from August 13 to September 27, 2009. Parties running for election. | | | | Unstable MAE depending on time window but larger than MAE reported by Tumasjan et al. 2010. Incorrect prediction when taking into account all parties running for election. |

Table II. Different studies on the feasibility of predicting elections with Twitter data characterized according to the scheme proposed above. Reports are ordered according to date of election and not of publication. Results appear on the right; those positive are shaded.

| | | | | | | |
|---|---|---|---|---|---|---|
| Metaxas et al. 2011 | US Senate special election in MA, 2010 (Jan. 19, 2010) | January 13 to 20, 2010. Candidate names. | No cleansing at all. | Number of tweets. Aggregated results at state level. Vote rates. | Winner prediction and MAE | Incorrect prediction. MAE 6.3%. |
| | | | | Lexicon-based sentiment analysis and vote share. Aggregated results at state level. Vote rates. | | Correct prediction. MAE 1.2% |
| | US elections in CO, 2010 (Nov. 2, 2010) | October 26 to November 1, 2010. Candidate names used to filter a *gardenhose* dataset. | | Number of tweets. Aggregated results at state level. Vote rates. | | Incorrect prediction. MAE 24.6% |
| | | | | Lexicon-based sentiment analysis and vote share. Aggregated results at state level. Vote rates. | | Correct prediction. MAE 12.4% |
| | US elections in NV, 2010 (Nov. 2, 2010) | | | Number of tweets. Aggregated results at state level. Vote rates. | | Correct prediction MAE 2.1% |
| | | | | Lexicon-based sentiment analysis and vote share. Aggregated results at state level. Vote rates. | | Incorrect prediction MAE 4.7% |

Table II (continuation). Results by [Metaxas et al. 2011] when trying to replicate results using methods analogous to those by [O'Connor et al. 2010; Tumasjan et al. 2010]. As it can be seen, results are inconclusive and there is no clear relation between MAE and accuracy of prediction.

| Metaxas et al. 2011 | US elections in CA, 2010 (Nov. 2, 2010) | October 26 to November 1, 2010. Candidate names used to filter a *gardenhose* dataset. | No cleansing at all. | Number of tweets. Aggregated results at state level. Vote rates. | Winner prediction and MAE | Correct prediction. MAE 3.8% |
|---|---|---|---|---|---|---|
| | | | | Lexicon-based sentiment analysis and vote share. Aggregated results at state level. Vote rates. | | Incorrect prediction. MAE 6.3% |
| | US elections in KY, 2010 (Nov. 2, 2010) | | | Number of tweets. Aggregated results at state level. Vote rates. | | Correct prediction. MAE 39.6% |
| | | | | Lexicon-based sentiment analysis and vote share. Aggregated results at state level. Vote rates. | | Correct prediction. 1.2% |
| | US elections in DE, 2010 (Nov. 2, 2010) | | | Number of tweets. Aggregated results at state level. Vote rates. | | Incorrect prediction. MAE 26.5% |
| | | | | Lexicon-based sentiment analysis and vote share. Aggregated results at state level. Vote rates. | | Incorrect prediction MAE 19.8% |

Table II (continuation). Results obtained by [Metaxas et al. 2011] when trying to replicate results using methods analogous to those by [O'Connor et al. 2010; Tumasjan et al. 2010]. As it can be seen, results are inconclusive and there is no clear relation between MAE and accuracy of prediction.

| | | | | | | |
|---|---|---|---|---|---|---|
| Livne et al. 2011 | US elections, 2010 (Nov. 2, 2010) | March 25, 2007 to November 1, 2010. Tweets and social graph for 700 candidates. | Not applicable. This method did not employ potential voter tweets but candidate data. | Regression models for binary results of races which included external data. Aggregated results at state level. Winner prediction. | Winner prediction | 81.5% accuracy when using external data alone. 83.8% accuracy when incorporating tweets (but not graph data). Not noticeable improvement. |
| Bermingham & Smeaton, 2011 | Irish general election, 2011 (Feb. 25, 2011) | February 8 to 25, 2011. Major parties. | No cleansing at all. | Number of tweets (different samples tested). Aggregated results at national level. Vote rates. | MAE against actual electoral results. | MAE 5.58% (uncompetitive with traditional polls). |
| | | | | ML-based sentiment analysis. Aggregated results at national level. Vote rates. | | MAE 3.67% (uncompetitive with traditional polls even after overfitting for using poll data for training). |
| Skoric et al., 2012 | Singaporean general election, 2011 (May 7, 2011) | April 1 to May 7, 2011. Tweets by 13,000 Singaporean political engaged users. Parties and candidates names were used to filter the tweets. | Only data produced by users located at Singapore was used. | Number of tweets. Aggregated results at national level. Vote rates. | | MAE 5.23% Inconclusive since pre-electoral polls are banned in Singapore. |

Table II (continuation). Studies by [Livne et al. 2011; Bermingham & Smeaton 2011; and Skoric et al. 2012]. The former is not properly a prediction method.

| | | | | | | |
|---|---|---|---|---|---|---|
| Tjong Kim Sang & Bos, 2012 | Dutch senate election, 2011 (May 23, 2011) | February 23 to March 1, 2011. Major parties. | No cleansing at all. | Number of tweets. Aggregated results at national level. Number of Senate seats. | Offset in number of seats as compared with actual results. | MAE (computed by this author) 1.33% Competitive with traditional polls. |
| | | | Attempt to debias data according to political leaning by using pre-electoral polling data | Sentiment analysis. Aggregated results at national level. Number of Senate seats. | | MAE (computed by this author) 2% Comparable to traditional polls although larger. |

Table II (ending). Study by [Tjong Kim Sang & Bos 2012]. MAE (computed by this author) is rather competitive with traditional polls. Interestingly, the more complex method incorporating sentiment analysis and some pre-election poll data underperforms the simpler relying on tweet counts.

## 2.3    Data cleansing

Data cleansing refers to measures adopted to make the prediction by (1) relying on those users who are prospective voters, (2) taking into account only those tweets dealing with the electoral process, and (3) correcting the collected data for any demographic bias in the Twitter user base. We will tackle with each of them individually.

The first one, denoted above as purity, involves those decisions adopted to select Twitter users that are likely voters in the election of interest. Needless to say, such information is unavailable and, hence, the most one can do is to limit the data collection to those users located in the area of interest.

This would discharge those users not providing a valid location or even emigrants eligible to vote, but it is a compromise solution to avoid taking into account users expressing their views on the campaign without being eligible to vote.

Such a solution is feasible by using just geolocated tweets or checking the location string of the users within the collection. Rather surprisingly, only two studies in the literature have applied such a measure. [Gayo-Avello 2011] by relying on tweets geolocated in counties of interest, and [Skoric et al. 2012] by limiting the dataset to those users located in Singapore.

Arguably, we could accept that [Tumasjan et al. 2010; Jungherr et al. 2011; Tjong Kim Sang & Bos 2012] guarantee to a certain extent purity of the collection on the basis of language use. The first two papers deal with German elections and, thus, it is assumed than tweets about German parties and politicians written in German are, very likely, produced by German users and not Austrian or Swiss users. In the same way, the third study deals with Dutch elections and the data is probably originated in Netherlands and not Belgium.

Nevertheless, when considering more globalized languages (as English, for instance) such an assumption is not acceptable and geolocation should be incorporated to guarantee the purity of the dataset.

The second measure to clean the data is denoising. This includes any post-processing of the dataset to remove tweets or users not dealing with the electoral process or not corresponding to prospective voters, respectively. In other words, it implies the removal of spam, rumors, propaganda, disinformation and users producing noisy tweets.

Table II reveals that no paper in the literature has adopted such measures although a few of them acknowledge the problem. For instance, [Metaxas et al. 2011] made this warning:

> "Spammers and propagandists write programs that create lots of fake accounts and use them to tweet intensively, amplifying their message, and polluting the data for any observer. It's known that this has happened in the past. It is reasonable that, if the image presented by social media is important to some (advertisers, spammers, propagandists), there will likely be people who will try to tamper with it."

What is more, they conducted a experiment to check the robustness of commonly applied sentiment analysis methods to such a kind of manipulation finding that:

> "[B]y just relying on polarity lexicons the subtleties of propaganda and disinformation are not only missed but even wrongly interpreted."

Hence, measures to filter noise in political tweets are not optional to produce accurate predictions.

The third and last data cleansing measure is debiasing. Twitter's user base is not a representative sample of the population, and that problem can be tackled with by determining the demographic strata users belong to and weighting their tweets accordingly.

The low representativeness of Twitter has been widely discussed. For instance, [danah boyd 2010] wrote the following:

> "Big Data presents new opportunities for understanding social practice. Of course the next statement must begin with a 'but.' And that 'but' is simple: […] just because you have a big N doesn't mean that it's representative or generalizable."

Besides, even in the United States Twitter use is minor (11% of Americans) and their users are "overwhelmingly young" [Lenhart and Fox 2009].

This low representativeness is a major problem because dominating demographic groups may tilt toward a few selected political options [Smith and Rainie 2008], and such a leaning heavily distort results [Gayo-Avello 2011].

Hence, the only way to fight this problem is by (1) determining as much demographic information about Twitter users as possible, and (2) weighting tweets from each group according to prior knowledge about their electoral involvement.

Needless to say, the first task is far from easy. Unlike other services such as Facebook, Twitter profiles do not include structured information. There is no way to indicate the user's sex or age and, instead, profiles consist of free text fields for name, location, website, and biography.

Nevertheless, it is not unsolvable and in a later section some references on this matter are commented. Indeed, Table II reveals that two papers have tried debiasing.

[Gayo-Avello 2011] was able to obtain the age for about 2,500 users in his dataset by crossing their full names and county with online public records. This way he found that the dataset was dominated by users in the 18-44 age interval. Then, by weighting their tweets according to age participation in the 2004 elections he was able to reduce the error from 13.10% to 11.61% –a significant boost in performance.

[Tjong Kim Sang & Bos 2012] tried a different approach: debiasing the data according to the presumed political leaning of the population. Certainly, such a feature is extremely relevant, especially if a "shy-Tory" effect[1] is suspected. Unfortunately, their results were inconclusive since (1) the authors had to rely on pre-electoral polling data which could be seen as overfitting; and (2) the performance of the method when debiasing was no better than a simpler method based on tweet counts.

Therefore, debiasing Twitter data according to demographic features of the users seems not only unavoidable but also to positively affect performance.

## 2.4    Method of prediction

### 2.4.1    Inferring votes by counting tweets

Two main methods have been used to infer votes from tweets. The first one, originally proposed by [Tumasjan et al. 2010], consists of merely counting the tweets mentioning a given candidate or party. The larger the number of tweets, the larger the vote rate.

Such a method is appealing for many reasons: it is easy to implement, it can be applied in near real-time, and it can be used both to obtain aggregated vote rates and to infer voting intentions for individuals (i.e. the candidate a user is mentioning the most would be his or her chose). Moreover, [Tumasjan et al. 2010] claimed the method exhibited good performance:

"The mere number of tweets reflects voter preferences and comes close to traditional election polls."

Indeed, Table II shows they reported an error of 1.65% for the German federal elections in 2009.

[Jungherr et al. 2011] later criticized some of the decisions taken by [Tumasjan et al. 2010], especially those regarding the selection of parties, and the period of data collection.

As it has been discussed, the selected time window has an impact but error values found by [Jungherr et al. 2011] were in the order of that reported by [Tumasjan et al. 2010]. Nevertheless, using a time window ending at the election day (the most plausible decision) produced an error of 2.13% which is substantially larger than both the original report of [Tumasjan et al. 2010] and traditional polls.

However, we have already stated that further research is needed with regards to periods of data collection and, moreover, there are no reasons to assume that tweet counting can be more sensitive to this issue than sentiment analysis.

Arguably, it is the selection of the candidates and parties to monitor the key aspect of this method. However, except for the analysis of the Pirate Party case by [Jungherr et al. 2011] there are no studies in this regard: all of the papers applying tweet counting relied on major parties.

Thus, are tweet counts for major parties a good predictor of voting intention? At first glance the answer could seem inconclusive; after all, half of the papers using that method correctly predicted the elections. However, from an intuitive point of view, it seems too good to be true. Does it mean that polarity in opinions does not matter?

In this regard, a experiment conducted by [Gayo-Avello 2011] is rather instructive. He compared the performance of both the tweet counting method and the lexicon-based sentiment analyzer against a random classifier.

---

[1] This effect refers to those conservative voters not disclosing their intentions in polls and, thus, biasing the corresponding predictions for conservative vote

To that end, he collected data from an informal opinion-poll conducted during the US presidential election in Twitter. A website called TwitVote[2] asked users to declare their votes with a tweet tagged with the hashtag *#twitvote*. By collecting those tweets published in election day, he was able to find the actual votes for a number of users. Besides, by taking into account the proportion of twitvotes for both candidates the performance for a random classifier could be computed and taken as a baseline.

This way, he found that tweet counts underperformed the random classifier for both candidates: slightly for Obama and by a huge margin for McCain. The lexicon-based sentiment classifier, however, outperformed the random classifier for both candidates.

In other words, no matter how appealing it looks like, raw tweet counts is not close to taking random choices but worse. Hence, and spite of purportedly positive results, such a method should be avoided in the future.

### 2.4.2 Inferring votes with sentiment analysis

The other popular method to infer voting intentions from tweets is sentiment analysis. The name is misleading because despite the extensive research conducted in that field (cf. [Liu 2012]) virtually all of the studies on electoral prediction have relied on the simplest of methods.

Except for [Bermingham and Smeaton 2011], and [Tjong Kim Sang and Bos 2012] who applied machine learning to train their sentiment classifiers –with mixed results, it must be said; the rest of studies have relied on lexicons to determine the polarity of tweets.

[O'Connor et al. 2010] were the first using that method. They relied on the lexicon by [Wilson et al. 2005] which consists of a list of terms labeled as positive or negative. Thus, tweets can be scored one way or the other, or even assigned both scores.

As with the previous method this one is also appealing because of its simplicity. However, it is also unsatisfactory. O'Connor et al. already found many examples of incorrectly detected sentiment although they still argued that

"With a fairly large number of measurements, these errors will cancel out relative to the quantity we are interested in estimating, aggregate public opinion."

This, however, can be problematic if the classifier is producing different amounts of errors for each candidate.

In this regard, the experiment by [Gayo-Avello 2011] from previous subsection is pertinent again. He checked the performance of a lexicon-based classifier similar to that used by [O'Connor et al. 2010] finding that precision for Obama was rather high (88.8%) but extremely poor for McCain (17.7%). Hence, although the method is certainly a classifier (since it outperforms a random one) it is not balanced and, therefore, it is unrealistic to expect errors to cancel out when aggregating results.

Finally, [Metaxas et al. 2011] conducted additional experiments with lexicon-based methods confirming, firstly, that their performance is only slightly better than that of a random classifier; and secondly, that misleading information and propaganda are missed or wrongly interpreted as candidate support.

In short, polarity based methods employed up to date:
1.  Miss the subtleties of political language.
2.  Exhibit very poor performance and,
3.  Produce unbalanced results making unrealistic to accept that errors will cancel out when aggregating data.

Therefore, sentiment analysis remains as an open challenge in this field of research.

## 2.5    Performance evaluation

### 2.5.1    Evaluation measures

The final question regarding electoral prediction methods is how to evaluate them. Needless to say, the actual outcome of the elections is needed but, what does that mean? The vote rates for each candidate? The seats achieved in congress or senate? The winner of the election?

---

[2] http://twitvote.twitmarks.com/

Moreover, at which level should predictions be computed and evaluated? This is obvious for local elections but not that much for national elections which could be computed as an aggregated popular vote or at more fine-grained levels.

Current research has produced predictions mainly at national level [Tumasjan et al. 2010; Jungherr et al. 2010; Bermingham and Smeaton 2011; Skoric et al. 2012; Tjong Kim Sang and Bos 2012] with a few papers focusing at state level [Gayo-Avello 2011; Metaxas et al. 2011].

Those predictions have been evaluated against vote rates [Gayo-Avello 2011; Tumasjan et al. 2010; Jungherr et al. 2011; Metaxas et al. 2011; Bermingham and Smeaton 2011; Skoric et al. 2012]; against number of seats [Tjong Kim Sang & Bos 2012]; and also as dichotomous decisions [Metaxas et al. 2011].

When predicting vote rates MAE (Mean Absolute Error) is commonly used after [Tumasjan et al. 2010]. This measure allows researchers to compare their method's performance against that of pre-electoral polls.

Winner prediction is also appealing but it can be misleading since no details are provided about how far or close the prediction was from the actual results and greatly depends on the granularity of the prediction. For instance, [Gayo-Avello 2011] predicted a Obama victory which was correct but that included a victory in Texas, which was incorrect.

Moreover, results by [Metaxas et al. 2011] shown in Table II reveal that both tweet counts and sentiment analysis were able to correctly guess half of the races. However, as it will be discussed in the following subsection this is probably not a great performance.

Finally, MAE is highly dependent on each race and election. For instance, the Senate election in Kentucky was correctly predicted with a MAE of 39.6% while a MAE of 6.3% produced an incorrect prediction in California (see Table II).

### 2.5.2   What would be an appropriate baseline?

The proper question is not which measure to apply but against which baseline to compare performance. With regards to winner prediction [Metaxas et al. 2011] suggested that:

"Given that, historically, the incumbent candidate gets re-elected about 9 out of 10 times, the baseline for any competent predictor should be the incumbent re-election rate."

Under such assumption it is pretty clear that guessing 50% of the races is rather far from competent.

With regards to vote rate prediction this author is not aware of any baseline akin to the previous one. Nevertheless, it seems plausible to use the results of the immediately prior election as a prediction.

Certainly, this has got issues: e.g., new parties running for election or coalitions created or dismantled between elections. Still, it is simple and can provide an intuitive hint about how "hard" or "easy" to predict an election can be.

Hence, such a baseline has been used to produce the data in Table III and to discuss the performance of the prior methods when predicting different elections. Both the table and the discussion appear in the following subsection.

## 2.6    Can or cannot elections be predicted from Twitter data?

Table II describes 14 different attempts to predict elections based on Twitter data[3]. Only half of them were successful, one was the result reported by [Tumasjan et al. 2010] –strongly contested by [Jungherr et al. 2011]– and the rest correspond to predictions in the paper by Metaxas et al. who, in turn, were able to predict half of the races.

All of this looks close to mere chance and, moreover, reported values of MAE are not directly comparable neither between papers nor races in the same election.

Hence, a new baseline was suggested in previous section: assuming past vote rates would happen again. Therefore, any method underperforming the MAE of such a baseline should be considered unsuccessful.

Thus, performance measures were computed for each of the elections studied in the literature for (1) the proposed baseline, (2) Twitter predictions based on tweet counts, and (3) Twitter predictions based on sentiment analysis where available.

Fortunately enough, all of the papers, except for [Tjon Kim Sang and Bos 2012] directly provide such information. In the later case it was computed from the results reported by the authors (seats in the Senate).

---

[3] As it will be shown in the commented bibliography [O'Connor et al. 2010; Livne et al. 2011] were not properly predictions.

Table III shows those performance measures which can help us to draw some conclusions about the predictive power of currently applied methods.

### 2.6.1 Can tweet counts predict elections?

There are three reports where predictions based on tweet counts outperform the baseline: [Tumasjan et al. 2010; Bermingham & Smeaton 2011; Tjon Kim Sang & Bos 2012]. Other three reports reveal that such a method underperforms the baseline: [Gayo-Avello 2011; Metaxas et al. 2011; Skoric et al. 2012].

Again, we could be facing something due to pure chance. On top of that, [Jungherr et al. 2011] showed that (1) prior important decisions regarding which parties to consider are required; and (2) performance measured by MAE strongly depends on the time window employed.

Indeed, all the researchers who have replicated the method by [Tumasjan et al. 2010] discharged minor parties –even those with results underperforming the baseline.

So, in short, Twitter prediction based on tweet counts:

1. Is too dependent on arbitrary decisions such as the parties or candidates to be considered, or the selection of a period for collecting the data.
2. Its performance is too unstable and strongly dependent on such parameterizations, and
3. Considering the reported results as a whole it seems plausible that positive results could have been due to chance or, even, to unintentional data dredging due to post hoc analysis.

In the absence of further research showing that the method can consistently predict future results for a number of elections outperforming both the incumbency and the past-results baselines, we must conclude that there is no strong evidence to consider it a valid method of prediction.

### 2.6.2 Can Twitter sentiment predict elections?

It is unclear the impact that sentiment analysis has in Twitter-based predictions. The studies applying that technique are fewer than those counting tweets and the picture they convey is confusing to say the least.

According to [Gayo-Avello 2011] sentiment analysis is better than raw tweet counts but it still underperforms the baseline.

[Metaxas et al. 2011] reveal that it outperforms not only raw counts but also the baseline. However, they also showed that lexicon-based methods are close to random classifiers (a finding also exposed by [Gayo-Avello 2011]) and that the proportion of correctly guessed races is no better than chance.

Results in [Bermingham & Smeaton 2011; Tjon Kim Sang & Bos 2012] introduce even more confusion. The former found that sentiment analysis outperforms the baseline but their method was overfitted since they incorporated data from pre-electoral polls, hence, being inconclusive. Results by the second authors reveal that their most complex method (involving not only sentiment analysis but also political leaning information derived from pre-electoral polls) outperforms the baseline but is not better than raw tweet counts.

In short, results are contradictory. However, taking into consideration that even naïve sentiment analysis seems to outperform the baseline it is clear that further research is needed in that line. This work concludes with some recommendations in that sense.

Table III. Performance measured as MAE for a naïve baseline predicting past vote rates will happen again and the two different kinds of Twitter predictions (i.e. those based on tweet counts and those relying on sentiment analysis). Those methods outperforming the baseline appear shaded.

| Election | Baseline | Twitter prediction (tweet counts) | Twitter prediction (sentiment analysis) |
|---|---|---|---|
| US presidential election, 2008 | 5.86% (states analyzed in Gayo-Avello 2011) | 15.87% (computed for this paper from data by Gayo-Avello 2011) | 13.10% 11.61% when debiasing according to age (Gayo-Avello 2011) |
| German federal election, 2009 | 3.75% | 1.65% (Tumasjan *et al.* 2010) From 1.51% to 3.34% (Jungherr *et al.* 2011) | Not available |
| US elections, 2010 | 8.85% (states analyzed in Metaxas *et al.* 2011) | 17.12% (Metaxas *et al.* 2011) | 7.58% (Metaxas *et al.* 2011) |
| Irish general election, 2011 | 6.26% | 5.58% (Bermingham and Smeaton 2011) | 3.67% includes polling data (Bermingham and Smeaton 2011) |
| Singaporean general election, 2011 | 3.36% | 5.23% (Skoric *et al.* 2012) | Not available |
| Dutch senate election, 2011 | 2.38% | 0.89% raw counts 1.33% normalized counts (computed for this paper by the author from data by Tjong Kim Sang and Bos 2012) | 2% normalized counts plus debiasing (computed for this paper by the author from data by Tjon Kim Sang and Bos 2012) |

# 3  AN ANNOTATED BIBLIOGRAPHY

This section covers the papers on electoral prediction analyzed above and, additionally, works on related areas such as credibility, rumors, and Twitter demographics.

## 3.1  Electoral prediction from Twitter data

### 3.1.1  O'Connor et al. 2010

This is the earliest paper discussing the feasibility of using Twitter data as a substitute for traditional polls although it does not describe any prediction method.

They employed a subjectivity lexicon to determine a positive and a negative score for each tweet in their dataset. Then, raw numbers of positive and negative tweets are used to compute a sentiment score.

Using that method, sentiment time series were prepared for a number of topics (namely, consumer confidence, presidential approval, and US 2008 Presidential elections). Both consumer confidence and presidential approval polls exhibited correlation with Twitter sentiment data, but no correlation was found with electoral polls.

### 3.1.2  Tumasjan et al. 2010, Jungherr et al. 2011, and Tumasjan et al. 2011

The first paper started the whole line of research analyzed in this work and proposed MAE as a measure of performance. It has two distinct parts: In the first one LIWC (Linguistic Inquiry and Word Count) is used to perform an analysis of the tweets related to different parties running for the German 2009 Federal election. In the second part, the authors state that the mere count of tweets mentioning a party accurately reflected the election results with a performance close to that of actual polls.

That claim was rebutted by [Jungherr et al. 2011] who pointed out that the method required arbitrary choices (e.g. not taking into account all the parties running for the elections) and that its results depended on the selected time window.

[Tumasjan et al. 2011] tried to dispel those doubts. Unfortunately, their counterarguments are not compelling enough and, besides, they toned down their previous conclusions: saying that Twitter data is not to replace polls but to complement them; or stating that the prediction method was not their main contribution[4].

### 3.1.3    Metaxas et al. 2011.

With [Jungherr et al. 2011] this is one of the few papers casting doubts on the predictive powers of Twitter. After analyzing results from a number of elections, they concluded that Twitter data is slightly better than chance when predicting elections. Hence, they suggested the use of incumbency as a baseline.

They also described three necessary standards for any method claiming predictive power: (1) it should be an algorithm, (2) it should take into account the demographic bias in Twitter's user base, and (3) it should be "explainable", i.e. black-box approaches should be avoided.

### 3.1.4    Livne et al. 2011

This is not a prediction method since it does not rely on users' tweets but in those by candidates plus their social graph. They also incorporate additional information such as the party a candidate belongs to, or incumbency.

They claim 88% precision when incorporating Twitter data (both tweets and graph) versus 81% precision without such data; the improvement is not substantial although noticeable.

Finally, it must be noted that elections are modeled as binary processes so important information is missed (such as in tight elections, or scenarios with coalitions).

### 3.1.5    Bermingham and Smeaton 2011

This paper discusses different approaches to incorporate sentiment analysis to a predictive method. The method was put to test with the 2011 Irish General Election revealing it was not competitive against traditional polls.

### 3.1.6    Gayo-Avello 2011

This paper describes how different methods failed to predict the 2008 US Presidential Election since they predicted an Obama victory in every state, including Texas. The methods were those proposed in [Tumasjan et al. 2010; O'Connor et al. 2010] and a post-mortem on the reasons for such a failure is provided.

A number of problems are suggested: (1) The "file-drawer" effect; (2) Twitter data is biased and is unrepresentative; and (3) the sentiment analyzers commonly used are only slightly better than random classifiers.

### 3.1.7    Tjong Kim Sang and Bos 2012

In this paper Twitter data regarding the 2011 Dutch Senate elections was analyzed. They found that tweet counting is a bad predictor and that sentiment analysis can improve performance.

Nevertheless, performance is below that of traditional polls and, moreover, the method relies to some extent on that polling data to correct for demographic bias.

### 3.1.8    Skoric et al. 2012

Similarly to previous paper, this also shows that there is certain correlation between Twitter chatter and votes but not enough to make accurate predictions –they found performance was much worse than that reported in [Tumasjan et al. 2010].

They argue that Twitter can provide a somewhat reasonable glimpse on national results but it fails when focusing on local levels. Hence, in addition to the technical caveats they discuss additional problems such as democratic maturity of the country, competitiveness of the election, and media freedom.

_______________________

[4] Even when the phrase *"predicting elections with Twitter"* prominently appears in the title of that paper

### 3.2 Sources of bias in Twitter

#### 3.2.1 Mislove et al. 2011

This paper analyzes a sample of Twitter users in the US along three different axes, namely, geography, gender and race/ethnicity.

The methods applied are simple albeit compelling. All of the data was inferred from the user profiles: geographical information was obtained from the self-reported location; gender was determined using the first name and statistical data from the US Social Security Administration; and the last name and data from the US Census was used to derive race/ethnicity.

Cleary, such methods are prone to error but it is probably rather tolerable and the conclusions of the study are sensible: highly populated counties are overrepresented, users are predominantly male, and Twitter is a non-random sample with regards to race/ethnicity.

They concluded that post-hoc corrections based on the over- and under-representation of different groups could be applied to improve predictions based on Twitter data.

#### 3.2.2 Mustafaraj et al. 2011

This paper provides compelling evidence on the existence of two extremely different behaviors in social media: on one hand there is a minority of users producing most of the content (vocal minority) and on the other there is a majority of users who hardly produce any content (silent majority).

These two groups are clearly separated and the vocal minority behaves as a resonance chamber spreading information aligned with their own opinions. Thus, they suggest extreme caution when building predictive models based on social media.

### 3.3 Denoising Twitter data

#### 3.3.1 Mustafaraj and Metaxas 2010

This paper introduces the concept of "Twitter-bomb": the use of fake accounts in Twitter to spread disinformation by "bombing" targeted users who, in turn, would retweet the message achieving viral diffusion.

They describe a smear campaign orchestrated by a Republican group against Democrat candidate Martha Coakley and how it was detected and aborted.

#### 3.3.2 Ratkiewicz et al. 2011

This paper describes the Truthy project inspired by the previous paper. Truthy is a system to detect astroturf political campaigns either to simulate widespread support for a candidate or to spread disinformation. The system is described in detail and a number of cases and performance analysis are provided.

#### 3.3.3 Castillo et al. 2011, and Morris et al. 2012

The first paper is, to the best of my knowledge, the first one describing a method to separate credible from not credible tweets. It describes in detail which features to extract from the tweets to then train a classifier.

[Morris et al. 2010] did not develop an automatic method to filter tweets but they conducted a survey to find the features that make users to perceive a tweet as credible. Content alone was found to be not enough to assess truthfulness so users rely on additional heuristics. These can be manipulated by the authors of tweets and, therefore, can affect credibility perceptions.

## 4 CONCLUSIONS

This paper concludes enumerating the main weaknesses of current research regarding political prediction with Twitter data. Then a list of open challenges and some recommendations for future research are provided.

### 4.1 Weaknesses in current research

The previous analysis of the literature revealed that the predictive power of Twitter regarding elections has been overstated and, indeed, recent results are not comparable to those by seminal papers.

Some of the reported positive results can probably be attributed to chance or involuntary data dredging and, moreover, simple baselines achieve better performance in many cases.

This is not surprising given that approaches up to date suffer a number of weaknesses that should be avoided in the future:

1. All of them are post hoc analysis.
2. Performance must be compared against reasonable baselines.
3. Sentiment analysis is applied with naïveté. Commonly used methods are slightly better than random classifiers and fail to catch the subtleties of political discourse.
4. All of the tweets are assumed to be trustworthy when it is not the case.
5. Demographics bias is neglected even when it is well known that social media is not a random sample of the population.
6. Self-selection bias is simply ignored. People tweet on a voluntary basis and, therefore, data is produced by those politically active.

### 4.2 Open challenges

A number of different lines of work must be pursued before making credible claims on electoral prediction from Twitter data:

- The accuracy of sentiment analysis of political tweets must improve. Humor and sarcasm will play a major role.
- New metrics of trustworthiness must be established; including, but not limited to, detection of propaganda, disinformation and robots, or credibility checking.
- Basic research on Twitter demographics and automatic profiling of users with regards to demographic attributes is required.
- Basic research on user participation regarding politics and self-selection bias is needed.

### 4.3 Recommendations for future research

Unless we avoid the mentioned weaknesses while focusing on the core lines of research depicted above, the obtained forecasts will be of questionable value or, worst, incorrect most of the time. This would taint a field of research that, properly explored, could be potentially useful.

Therefore, a number of requirements are needed to guarantee not only the quality of the outcomes but also the generality of the methods applied. Such requirements are highly related to the characterization scheme depicted earlier:

1. Performance should be evaluated against sound and credible baselines.
2. The period and method of collection must be clearly specified and justified. Any prior decision regarding candidates or parties to be discharged must be justified.
3. Data purity should be guaranteed. That is, only tweets by users eligible to vote should be collected.
4. State of the art methods of sentiment analysis should be applied instead of simplistic and crude approaches.
5. Noise in the data should be kept to a minimum. That is, serious attempts to remove spam and disinformation should be made.
6. Bias in the data should be at least acknowledged and analyzed. Attempts to remove demographic bias are encouraged. Self-selection bias should be analyzed at least.

### 4.4 Final note

This paper must conclude, however, with a pessimistic remark. All previous considerations are a sine qua non to consider electoral predictions based on social media completely analogous to traditional polls.

Unfortunately, as we move down the list of requirements the tasks are harder and even unattainable. Improving sentiment analysis with regards to political tweets will be difficult but probably achievable. Fighting demographic bias is at least conceivable in principle. However, accounting for self-selection bias could be unviable on a general basis; at least, not using Twitter data alone.

Hence, social media may very well provide a glimpse on the outcome of elections, and the best the methods the most accurate the glimpse. However, current state of the art does not provide any strong evidence to support the idea it is going to replace traditional polls in the short term.

# REFERENCES

ASUR, S., AND HUBERMAN, B.A. 2010, "Predicting the Future with Social Media", in Proceedings of the IEEE/WIC/ACM International Conference on Web Intelligence and Intelligent Agent Technology, IEEE Computer Society, Los Alamitos, CA, USA, 492–499.

BERMINGHAM, A., AND SMEATON, A. 2011, "On Using Twitter to Monitor Political Sentiment and Predict Election Results", paper presented at the Workshop on Sentiment Analysis where AI meets Psychology, November 13, 2011, Chiang Mai, Thailand.

BOLLEN, J., MAO, H., ZENG, X.J. 2011, "Twitter mood predicts the stock market", J. Comput. Science, vol. 2, no. 1, 1–8.

BOYD, D. 2010, "Big Data: Opportunities for Computational and Social Sciences", April 17, 2010, available at: http://www.zephoria.org/thoughts/archives/2010/04/17/big-data-opportunities-for-computational-and-social-sciences.html (accessed 14 June 2012).

CASTILLO, C., MENDOZA, M., AND POBLETE, B. 2011, "Information credibility on twitter", in Sadagopan, S. et al. (Eds.) Proceedings of the 20th international conference on World Wide Web, ACM, New York, NY, USA, pp. 675–684.

CHOI, H., AND VARIAN, H. (2009a), "Predicting the Present with Google Trends", technical report, Google Inc. April 10, 2009, available at: http://google.com/googleblogs/pdfs/google_predicting_the_present.pdf (accessed 14 June 2012).

CHOI, H., AND VARIAN, H. (2009b), "Predicting Initial Claims for Unemployment Benefits", technical report, Google Inc. July 5, 2009, available at: http://research.google.com/archive/papers/initialclaimsUS.pdf (accessed 14 June 2012).

FANELLI, D. 2010, "Do Pressures to Publish Increase Scientists' Bias? An Empirical Support from US States Data", PLoS ONE, vol. 5, no. 4, available at: http://www.plosone.org/article/info:doi/10.1371/journal.pone.0010271 (accessed 14 June 2012).

GAYO-AVELLO, D. 2011, "Don't turn social media into another 'Literary Digest' poll", Communications of the ACM, vol. 54, no. 10, 121–128.

GINSBER, J., MOHEBBI, M.H., PATEL, R.S., BRAMMER, L., SMOLINSKI, M.S., AND BRILLIANT, L. 2009, "Detecting influenza epidemics using search engine query data", Nature, vol. 457, 1012–1014.

GRUHL, D., GUHA, R., KUMAR, R., NOVAK, J., AND TOMKINS, A. 2005, "The predictive power of online chatter", in Grossman, R. et al. (Eds.), Proceedings of the eleventh ACM SIGKDD international conference on Knowledge discovery in data mining, ACM, New York, NY, USA, 78–87.

JUNGHERR, A., JÜRGENS, P., AND SCHOEN, H. 2011, "Why the Pirate Party won the German election of 2009 or the trouble with predictions: A response to Tumasjan, A., Sprenger, T.O., Sander, P.G., & Welpe, I.M. "Predicting elections with Twitter: What 140 characters reveal about political sentiment", Social Science Computer Review, published online before print April 25, 2011, available at: http://ssc.sagepub.com/content/early/2011/04/05/0894439311404119.abstract (accessed 14 June 2012).

LENHART, A., AND FOX, S. 2009, "Twitter and status updating", report, Pew Internet and American Life, February 12 2009, available at: http://www.pewinternet.org/Reports/2009/Twitter-and-status-updating.aspx (accessed 14 June 2012).

LIU, B. 2012, Sentiment Analysis and Opinion Mining, Morgan & Claypool Publishers.

LIVNE, A., SIMMONS, M.P., ADAR, E., AND ADAMIC, L.A. 2011, "The Party Is Over Here: Structure and Content in the 2010 Election", paper presented at the 5th International AAAI Conference on Weblogs and Social Media, July 17-21, 2011, Barcelona, Spain.

METAXAS, P.T., MUSTAFARAJ, E., AND GAYO-AVELLO, D. 2011, "How (Not) to Predict Elections", in Proceedings of PASSAT/SocialCom 2011, 2011 IEEE Third International Conference on Privacy, Security, Risk and Trust (PASSAT) and 2011 IEEE Third International Conference on Social Computing (SocialCom), IEEE Computer Society, Los Alamitos, CA, USA, 165–171.

MING FAI WONG, F., SEN, S., AND CHIANG, M. 2012, "Whay Watching Movie Tweets Won't Tell the Whole Story?", arXiv preprint, Princeton University, March 21, 2012, available at: http://arxiv.org/abs/1203.4642 (accessed 14 June 2012).

MISHNE, G., AND DE RIJKE, M. 2006, "Capturing global mood levels using blog posts", paper presented at AAAI 2006 Spring Symposium on Computational Approaches to Analysing Weblogs, March 27-29, 2006, California, USA.

MISHNE, G., AND GLANCE, N. 2006, "Predicting movie sales from blogger sentiment", paper presented at AAAI 2006 Spring Symposium on Computational Approaches to Analysing Weblogs, March 27-29, 2006, California, USA.

MISLOVE, A., LEHMANN, S., AHN, Y.Y., ONNELA, J.P., AND ROSENQUIST, J.N. 2011, "Understanding the Demographics of Twitter Users", paper presented at the 5th International AAAI Conference on Weblogs and Social Media, July 17-21, 2011, Barcelona, Spain.

MORRIS, M.R., COUNTS, S., ROSEWAY, A., HOFF, A., AND SCHWARZ, J. 2012, "Tweeting is believing?: understanding microblog credibility perceptions", in Poltrock, S. et al. (Eds.) Proceedings of the ACM 2012 conference on Computer Supported Cooperative Work, ACM, New York, NY, USA, 441–450.

MUSTAFARAJ, E., AND METAXAS, P.T. 2010, "From Obscurity to Prominence in Minutes: Political Speech and Real-Time Search", paper presented at WebSci10: Extending the Frontiers of Society On-Line, April 26-27, 2010, Raleigh, NC, USA.

MUSTAFARAJ, E., FINN, S., WHITLOCK, C., AND METAXAS, P.T. 2011, "Vocal Minority Versus Silent Majority: Discovering the Opionions of the Long Tail", in Proceedings of PASSAT/SocialCom 2011, 2011 IEEE Third International Conference on Privacy, Security, Risk and Trust (PASSAT) and 2011 IEEE Third International Conference on Social Computing (SocialCom), IEEE Computer Society, Los Alamitos, CA, USA, 103–110.

O'CONNOR, B., BALASUBRAMANYAN, R., ROUTLEDGE, B.R., AND SMITH, N.A. 2010, "From Tweets to Polls: Linking Text Sentiment to Public Opinion Time Series", paper presented at the 4th International AAAI Conference on Weblogs and Social Media, May 23-26, 2010, Washington, USA.

RATKIEWICZ, J., CONOVER, M., MEISS, M., GONÇALVES, B., FLAMMINI, A., AND MENCZER, F. 2011, "Detecting and Tracking Political Abuse in Social Media", paper presented at the 5th International AAAI Conference on Weblogs and Social Media, July 17-21, 2011, Barcelona, Spain.

SKORIC, M., POOR, N., ACHANANUPARP, PL, LIM, E.P., AND JIANG, J. 2012, "Tweets and Votes: A Study of the 2011 Singapore General Election", in Proceedings of 45th Hawaii International International Conference on Systems Science (HICSS-45 2012), IEEE Computer Society, Los Alamitos, CA, USA, 2583–2591.

SMITH, A., AND RAINIE, L. 2008, "The Internet and the 2008 election", report, Pew Internet and American Life, June 15, 2008, available at: http://www.pewinternet.org/Reports/2008/The-Internet-and-the-2008-Election.aspx (accessed 14 June 2012).

TJONG KIM SANG, E., AND BOS, J. 2012, "Predicting the 2011 Dutch Senate Election Results with Twitter", paper presented at the 13th Conference of the European Chapter of the Association for Computational Linguistics, April 23-27, 2012. Avignon, France.

TUMASJAN, A., SPRENGER, T.O., AND WELPE, I.M. 2010, "Predicting Elections with Twitter: What 140 Characters Reveal about Political Sentiment", paper presented at the 4th International AAAI Conference on Weblogs and Social Media, May 23-26, 2010, Washington, USA.

TUMASJAN, A., SPRENGER, T.O., SANDNER, P.G., AND WELPE, I.M. 2011, "Where There is a Sea There are Pirates: Response to Jungherr, Jürgens, and Schoen", Social Science Computer Review, published online before print May 18, 2011, available at: http://ssc.sagepub.com/content/30/2/235.abstract (accessed 14 June 2012).

WILSON, T., WIEBE, J., AND HOFFMANN, P. 2005, "Recognizing Contextual Polarity in Phrase-Level Sentiment Analysis", paper presented at the Human Language Technology Conference and Conference on Empirical Methods in Natural Language Processing, October 6-8, 2005, Vancouver, B.C., Canada.